\documentclass[conference, 10pt]{IEEEtran}

\usepackage[normalem]{ulem}

\usepackage{cite}
\usepackage[linesnumbered, algo2e, ruled,norelsize]{algorithm2e}
\usepackage{amsmath,amssymb,amsfonts}
\usepackage{algorithmic}
\usepackage{graphicx}
\usepackage{textcomp}
\usepackage{graphicx}
\usepackage{dsfont}
\usepackage{caption}
\usepackage{subcaption}
\usepackage{latexsym}
\usepackage{amsmath}
\usepackage[mathscr]{euscript}
\usepackage[linesnumbered, algo2e, ruled,norelsize]{algorithm2e}
\SetArgSty{textnormal}
\usepackage{threeparttable}
\usepackage{threeparttablex}
\usepackage{commath}
\usepackage{mathtools}
\usepackage{slashbox}
\usepackage{pict2e}
\usepackage{epstopdf}
\usepackage{verbatim}
\usepackage{subfloat}
\usepackage{tabu}
\usepackage{booktabs}
\usepackage{float}
\usepackage{amssymb}
\usepackage{url}
\usepackage{multirow}
\usepackage{graphicx}
\usepackage{mathptmx}
\usepackage{cite}
\usepackage{footnote}
\usepackage{array}
\usepackage[table]{xcolor}
\usepackage{tabularx}
\usepackage{ulem}
\usepackage{array}
\usepackage{slashbox}
\usepackage{lipsum}
\usepackage{stfloats}
\newcolumntype{P}[1]{>{\centering\arraybackslash}p{#1}}

\usepackage{amsthm} 
\makeatletter
\newcommand{\leftlabel}[1]{&&
  \refstepcounter{equation}\ltx@label{#1}%
  \tagform@{\theequation}&&}
\makeatletter

\def\BibTeX{{\rm B\kern-.05em{\sc i\kern-.025em b}\kern-.08em
    T\kern-.1667em\lower.7ex\hbox{E}\kern-.125emX}}

\begin{document}

\title{Design of Artificial Noise for Physical Layer Security in Visible Light Systems with Clipping}

\author{\IEEEauthorblockN{Thanh V. Pham\IEEEauthorrefmark{1},
Steve Hranilovic\IEEEauthorrefmark{2}, and
Susumu Ishihara\IEEEauthorrefmark{1}}
\IEEEauthorblockA{\IEEEauthorrefmark{1}Department of Mathematical and Systems Engineering, Shizuoka University, Shizuoka, Japan.}
\IEEEauthorblockA{\IEEEauthorrefmark{2}Department of Electrical and Computer Engineering, McMaster University, ON, Canada.}
Emails: \IEEEauthorrefmark{1}\{pham.van.thanh, ishihara.susumu\}@shizuoka.ac.jp,  \IEEEauthorrefmark{2}hranilovic@mcmaster.ca
}
\maketitle
\begin{abstract}
Though visible light communication (VLC) systems are contained in a given room, ensuring their security amongst users in a room is essential. In this paper, the design of artificial noise (AN)  to enhance physical layer security in VLC systems is studied in the context of input signals with no explicit amplitude constraint (such as multicarrier systems).  In such systems, clipping is needed to constrain the input signals within the limited linear ranges of the LEDs.  However, this clipping process gives rise to non-linear clipping distortion, which must be incorporated into the AN design. To facilitate the solution of this problem, a sub-optimal design approach is presented using the Charnes-Cooper transformation and the convex-concave procedure (CCP). Numerical results show that the clipping distortion significantly reduces the secrecy level, and using AN is advantageous over the no-AN scheme in improving the secrecy performance.         
\end{abstract}



\section{Introduction}
Visible light communications (VLC) is a promising wireless technology to complement existing radio-based wireless communication systems \cite{pathak2015}. Research and development of VLC systems have been motivated due to the increasing need for high-speed wireless connections, the scarcity of radio frequency (RF) spectrum, and the popularity of using light-emitting diodes (LEDs) for illumination. 

As visible light is confined by opaque objects, VLC systems are expected to be more secure than their RF counterparts. Nonetheless, there is still a security risk within the illumination areas of VLC systems due to the broadcast nature of the visible light signal. In this regard, over the past few years, there has been an increasing interest in applying physical layer security (PLS) to enhance message confidentiality in VLC systems \cite{arfaoui2020}.
Fundamental analyses on the secrecy capacity of systems consisting of a single transmitter, single legitimate user, and single eavesdropper were presented in \cite{wang2018physical}, where lower and upper secrecy capacity bounds were derived under an average optical intensity constraint, and  both average and peak optical intensity constraints. In indoor scenarios, multiple LED luminaries are often deployed to provide sufficient illumination. As a result, multi-transmitter VLC systems are more relevant in practice \cite{zeng2009high}. 

The use of multiple transmitter luminaires offers spatial degrees of freedom that can be exploited to improve the secrecy performance by means of precoding \cite{pham2017secrecy,Cho2020} and artificial noise (AN). In essence, AN is a jamming signal which is purposely generated by the transmitters to degrade the quality of the eavesdropper channel while introducing little or no interference to the legitimate user channel. 
The use of AN in improving the secrecy performance of VLC systems has been studied intensively in the literature \cite{Shen2016,Pham2018,Cho2018,Cho2019,Cho2021,pham2020energy}. 


It is well-known that LEDs have specific input ranges over which the emitted optical power is linearly related to the amplitude of the input drive current. Due to this characteristic,  previous work assumed that the AN and information-bearing signals are both amplitude-constrained  to ensure that the amplitude of the LED input drive current can be properly constrained within the LED linear range. However, the assumption of amplitude-constrained signals does not apply to multicarrier signals such as orthogonal frequency division multiplexing (OFDM), which are expected to be widely employed for high-speed VLC systems \cite{Serafimovski2021}. Multicarrier signals can often have  large peak-to-average power ratios. Thus clipping is unavoidable in VLC OFDM systems and it is necessary to constrain the signals within the LED linear range.  For the design of AN for VLC OFDM systems, the  nonlinear clipping distortion introduced renders  optimal design extremely difficult. Hence, in this paper a sub-optimal design is presented  which uses the Charnes-Cooper transformation and the convex-concave procedure (CCP). Simulation results show that the clipping distortion has a significant impact on the secrecy level, and using AN can considerably improve the secrecy performance. 

The balance of the paper is organized as follows. Sec. \ref{system} outlines the system and channel models. The AN transmission scheme and design approach are then described. Numerical results are discussed in Sec. \ref{result}. Finally, Sec. \ref{conclusion} concludes the paper. 

{\it{Notation}}: The following notations are used throughout the paper. Bold uppercase and lowercase letters (e.g., $\mathbf{H}$ and $\mathbf{h}$) represent matrices and vectors, respectively. $[\mathbf{H}]_{m, n}$ indicate the element at the $m$-th row and $n$-th column of $\mathbf{H}$ while $[\mathbf{h}]_n$ indicate the $n$-th element of $\mathbf{h}$. Also, $\mathbb{E}[\cdot]$ is the expected value operation, $\lVert\cdot\rVert$ is the Euclidean norm, and $\odot$ is the element-wise product operation. 

\section{System and AN Transmission Models}
\label{system}
\begin{figure}[ht]
    \centering
    \includegraphics[width = 8.0 cm, height = 5.5cm]{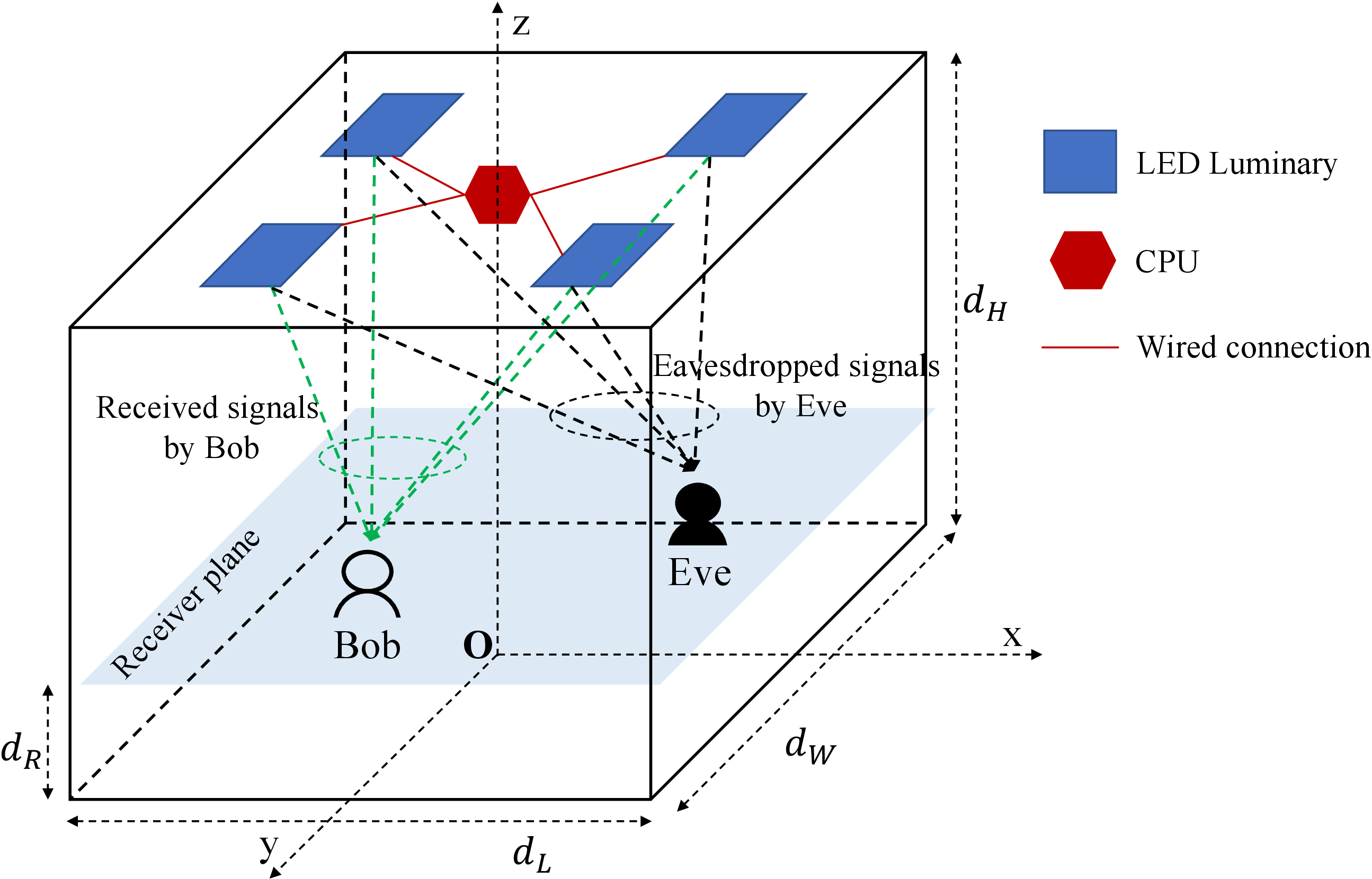}
    \caption{System Configuration.}
    \label{Fig0}
\end{figure}
Consider a multi-transmitter VLC system that consists of a legitimate user (Bob) and an eavesdropper (Eve) coexisting in a room of dimension $d_L~\text{(m)} \times d_W~\text{(m)} \times d_H~\text{(m)}$ as illustrated in Fig.~\ref{Fig0}. There are $N_T$ LED luminaries connected to a Central Processing Unit (CPU) through wired connections. Assume that Bob and Eve are both equipped with a receiver having a single photodiode (PD), and all receivers are on the same receiver plane which is $d_R~\text{(m)}$ above the floor. For the sake of simplicity, only the line-of-sight (LoS) propagation channel is considered. Let $\mathbf{h}_{\text{B}} = \begin{bmatrix}h_{1,\text{B}} & h_{2,\text{B}} & \hdots & h_{N_T, \text{B}}\end{bmatrix}^T$ be the channel vector of Bob where $h_{k, \text{B}}$ is the channel gain from the $k$-th LED luminary\footnote{The LoS channel gain model in the case of Lambertian emission pattern can be found in \cite{arfaoui2020} and references therein.}. 
Similarly, denote $\mathbf{h}_{\text{E}} = \begin{bmatrix}h_{1, \text{E}} & h_{2, \text{E}} & \cdots & h_{N_T, \text{E}}\end{bmatrix}^T$ as the channel vector of Eve. 
\subsection{AN transmission scheme}
The IEEE 802.11bb standard advocates for the use of DC-biased optical (DCO) OFDM to support high data rates (up to 10 Gbps) transmission in VLC \cite{Serafimovski2021}. Unlike PAM and OOK signaling considered in previous studies, OFDM signals are inherently not subject to amplitude constraint and can have multiple high peaks. Hence, clipping is necessary to fit the signals into the LED linear range. 

Let $d$ and $z$ be the information-bearing and AN signals, respectively. In previous studies, $d$ and $z$ were either assumed to follow the uniform distribution \cite{Pham2018,pham2020energy} or the truncated Gaussian distribution \cite{Zaid2015,Cho2019, Arfaoui2019} over the normalized amplitude range $[-1~1]$. In the case of OFDM, when a large number of sub-carriers is used (e.g., 64 or more), the signals can be well approximated to be Gaussian and have domain over the real numbers. Hence, for the sake of analysis, let $d$ and $z$ be time-domain Gaussian information-bearing and AN signals. Without loss of generality, assume that $d$ and $z$ are independent zero-mean and have unit variance. At the $n$-th luminaire, denote $v_n$ and $w_n$ as the precoders of $d$ and $z$, respectively. The combined information-bearing and AN current is given by
\begin{align}
    x_n = v_nd + w_nz.
\end{align}
\begin{figure}[ht]
    \centering
    \includegraphics[width = 8.0 cm, height = 5.5cm]{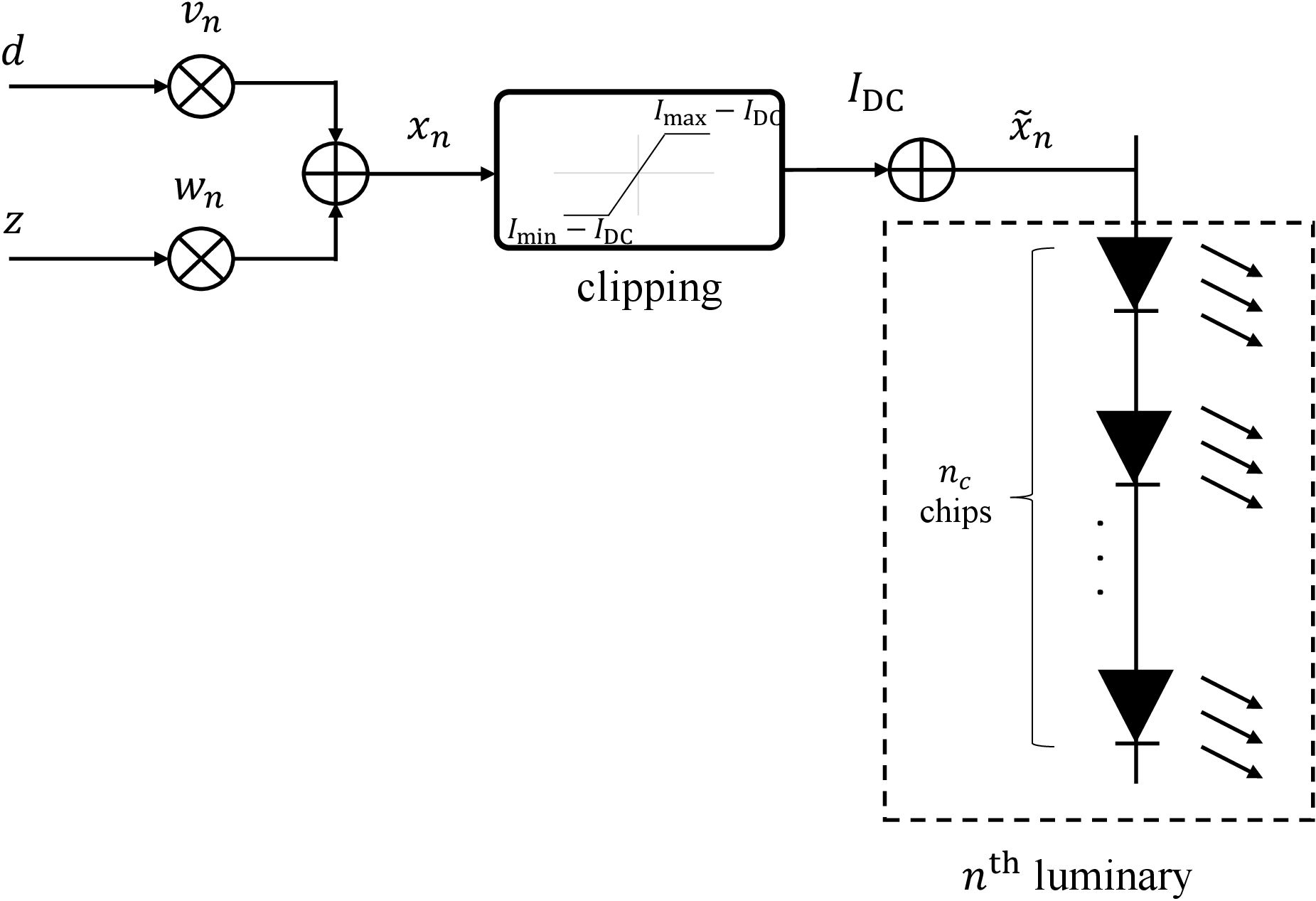}
    \caption{AN transmission scheme for non-amplitude constrained signals.}
    \label{Fig3}
\end{figure}
Since $d$ and $z$ are not amplitude-constrained, $x_n$ must be clipped to limit its amplitude within the LED operational range, which is assumed to be in the range $[I_{\text{min}},~ I_{\text{max}}]$. Since $d\sim \mathcal{N}(0, 1)$ and $z \sim \mathcal{N}(0, 1)$, it follows that $x_n \sim \mathcal{N}\left(0, v^2_n + w^2_n\right)$.
As illustrated in Fig.~\ref{Fig3}, a nonlinear clipping is applied to $x_n$ at a bottom level $I_{\text{min}}-I_{\text{DC}}$ and at a top level $I_{\text{max}}-I_{\text{DC}}$, where $I_{\text{DC}}$ is the DC-bias current which is added to the clipped signal. Thus, the resulting distorted signal $\Tilde{x}_n$ can be modeled using the Bussgang theorem as  
\begin{align}
    \Tilde{x}_n = R_n\left(v_nd + w_nz\right) + I_{\text{DC}} + \zeta_n,
\end{align}
where $R_n \in [0,~1]$ is the attenuation factor and $\zeta_n$ is the clipping noise \cite{Dimitrov2012}. An increased severity of clipping distortion is characterized by a small value of $R_n$ and a large variance of $\zeta_n$. If we denote $\alpha_n = \frac{I_{\text{min}} - I_{\text{DC}}}{\sqrt{v^2_n + w^2_n}}$ and $\beta_n = \frac{I_{\text{max}}- I_{\text{DC}}}{\sqrt{v^2_n + w^2_n}}$ then $R_n = Q(\alpha_n) - Q(\beta_n)$. The clipping noise $\zeta_n$ can be approximately modeled as a zero-mean Gaussian random variable whose variance is given by
\begin{align}
    \sigma^2_{\text{clip}, n} & \nonumber \\ =   &\Big(R_n+\alpha_n\phi(\alpha_n)-\beta_n\phi(\beta_n)  +\alpha_n^2(1-Q(\alpha_n))+\beta_n^2Q(\beta_n) \big. \nonumber \\ & \left. -\big(\phi(\alpha_n) - \phi(\beta_n) + (1-Q(\alpha_n))\alpha_n + Q(\beta_n)\beta_n\big)^2 \right. \!\!\! -  R_n^2\Big) \nonumber \\ & \times \left(v^2_n + w^2_n\right),
\end{align}
where $\phi(t) = \frac{1}{2\pi}\exp\left(\frac{-t^2}{2}\right)$ and $Q(t) = \frac{1}{2\pi}\int_{t}^{\infty}\exp\left(\frac{-u^2}{2}\right)\text{d}u$.

Suppose that there are $n_c$ closely packed LED chips in each luminaire. The instantaneous emitted optical power from the $n$-th LED luminary is then
\begin{align}
    P_{n, \text{optical}} = n_c\eta\left(R_n\left(v_nd + w_nz\right) + I_{\text{DC}} + \zeta_n\right),
\end{align}
where $\eta$ is the LED conversion factor. Since the LED chips in a given luminaire are assumed to be closely packed, it is reasonable to assume that the channel gains from them to a receiver are identical. The output electrical signal for Bob from all luminaires in the room and after the removal of the DC bias current is
\begin{align}
    r_{\text{B}} = \gamma\eta n_c\left(\left(\mathbf{h}_{\text{B}}\odot\mathbf{R}\right)^T\mathbf{v}d  + \left(\mathbf{h}_{\text{B}}\odot\mathbf{R}\right)^T\mathbf{w}z + \mathbf{h}^T_{\text{B}}\pmb{\zeta}\right) + n_{\text{B}},
    \label{user-received-signal}
\end{align}
where $\mathbf{R} = \begin{bmatrix}R_1 & R_2 & \cdots & R_{N_T}\end{bmatrix}^T$, $\mathbf{v} = \begin{bmatrix}v_1 & v_2 & \cdots & v_{N_T}\end{bmatrix}^T$, $\mathbf{w} = \begin{bmatrix}w_1 & w_2 & \cdots & w_{N_T}\end{bmatrix}^T$, and $\pmb{\zeta} = \begin{bmatrix}\zeta_1 & \zeta_2 & \cdots & \zeta_{N_T}\end{bmatrix}^T$. The receiver noise $n_{\text{B}}$ can be well approximated being Gaussian distributed with zero mean and variance $\sigma^2_{\text{B}}$ given by
\begin{align}
    \sigma^2_{\text{B}} = 2e\gamma\overline{P}_{r, \text{B}}B_{\text{mod}} + 4\pi e\gamma A_r\chi_{\text{amb}}\left(1 - \cos\Psi\right)B_{\text{mod}} + i^2_{\text{amp}}B_{\text{mod}},
    \label{BobReceiverNoise}
\end{align}
where $e$ is the elementary charge, $\gamma$ is the PD responsivity, $B_{\text{mod}}$ is the signal bandwidth, $\chi_{\text{amb}}$ is the ambient light photocurrent, and $i^2_{\text{amp}}$ is the pre-amplifier noise current density. $A_r$ and $\Psi$ are the active area and the optical field of view (FoV) of the PD, respectively. $\overline{P}_{r, \text{B}} = n_c\eta \sum_{n=1}^{N_T}h_{n, \text{B}}\mathbb{E}[\Tilde{x}_n]$ is the average received optical power at Bob's receiver. The signal-to-interference-plus-noise ratio (SINR) of Bob's received signal is then given by
\begin{align}
    \text{SINR}_{\text{B}} = \frac{\left(n_c \left(\mathbf{h}_{\text{B}}\odot\mathbf{R}\right)^T\mathbf{v}\right)^2}{ \left(n_c\left(\mathbf{h}_{\text{B}}\odot\mathbf{R}\right)^T\mathbf{w}\right)^2 + \left(n_c\mathbf{h}^T_{\text{B}}\pmb{\sigma}_{\text{clip}}\right)^2 + {\sigma}^2_{\text{B, norm}}},
    \label{BobSINR3}
\end{align}
where $\pmb{\sigma}_{\text{clip}} = \begin{bmatrix}\sigma_{\text{clip}, 1} & \sigma_{\text{clip}, 2}, & \hdots& \sigma_{\text{clip}, N_T}\end{bmatrix}^T$ and ${\sigma}^2_{\text{B, norm}} = \frac{\sigma^2_{\text{B}}}{\left(\gamma\eta\right)^2}$.

It is often unrealistic to assume that $\mathbf{h}_{\text{E}}$ is known at the transmitter due to the fact that Eve tends to hide itself from being detected. Nonetheless, the design of AN to provide an upper limit on secrecy performance when $\mathbf{h}_{\text{E}}$ is known is investigated here. 
Similar to Bob, the SINR of Eve's wire-taped signal is then given by
\begin{align}
    \text{SINR}_{\text{E}} = \frac{\left(n_c \left(\mathbf{h}_{\text{E}}\odot\mathbf{R}\right)^T\mathbf{v}\right)^2}{ \left(n_c\left(\mathbf{h}_{\text{E}}\odot\mathbf{R}\right)^T\mathbf{w}\right)^2 + \left(n_c\mathbf{h}^T_{\text{E}}\pmb{\sigma}_{\text{clip}}\right)^2 + {\sigma}^2_{\text{E, norm}}},
    \label{EveSINR3}
\end{align}
where $\sigma^2_{\text{E, norm}} = \frac{\sigma^2_{\text{E}}}{\left(\gamma\eta\right)^2}$.

\subsection{AN design} \label{sec:ANdesign}
Optimization problems in physical layer security often utilize the secrecy rate as the objective function. In addition to the secrecy rate, the secrecy performance can also be evaluated indirectly through the pair of SINRs of Bob's and Eve's received signal, which is often easier to handle. Moreover, by taking the SINR of Eve's received signal into the design, one can precisely (or qualitatively) control the performance of Eve' channel, such as the bit-error rate. In this work, we, therefore, study AN design problems using $\text{SINR}_{\text{B}}$ and $\text{SINR}_{\text{E}}$. Specifically,  $\text{SINR}_{\text{B}}$ is maximized while constraining $\text{SINR}_{\text{E}}$ to a predefined maximum threshold, denoted as $\lambda$. Additionally, the sum-power of the information-bearing and AN signals at the $n$-th luminary is constrained to a maximum allowable value of $P_n$. The AN design problem is thus given as follows
\begin{subequations}
\label{OptProb2}
    \begin{alignat}{2}
        &\underset{\mathbf{v}, \mathbf{w}}{\text{maximize}} & \hspace{2mm} & \text{SINR}_{\text{B}} \label{obj2}\\
        &\text{subject to }  &  & \nonumber \\
        & & & \text{SINR}_{\text{E}}  \leq \lambda, \label{constraint21}\\
        & & & \left[\mathbf{v}\right]^2_n + \left[\mathbf{w}\right]^2_n \leq P_n, ~~\forall n = 1, 2, ..., N_T \label{constraint22}.
    \end{alignat}
\end{subequations}
Note that, instead of \eqref{constraint22}, one can also apply a constraint on the sum-power of the information-bearing and AN signals over all luminaries, which is
\begin{align}
    \left\lVert\mathbf{v}\right\rVert^2 + \left\lVert\mathbf{w}\right\rVert^2 \leq P,
    \label{sumPowerConstraint}
\end{align}
where $P = \sum_{n = 1}^{N_T} P_n$. Compared with \eqref{sumPowerConstraint}, the advantage of using \eqref{constraint22} is two-fold. Firstly, it is clear that \eqref{constraint22} is a generalization of \eqref{sumPowerConstraint}. Secondly, using \eqref{constraint22} facilitates the handling of $\text{SINR}_{\text{B}}$ and $\text{SINR}_{\text{E}}$ as it can be seen that $\mathbf{R}$, $\pmb{\sigma}_{\text{clip}}$, $\sigma^2_{\text{B, norm}}$, and $\sigma^2_{\text{E, norm}}$ are functions of  $\left[\mathbf{v}\right]^2_n + \left[\mathbf{w}\right]^2_n$ at each individual luminary. Nonetheless, due to the complex non-linearity of $\mathbf{R}$, $\pmb{\sigma}_{\text{clip}}$, ${\sigma}^2_{\text{B, norm}}$, and ${{\sigma}}^2_{\text{E, norm}}$, it is extremely challenging (if not impossible) to optimally solve the above problem. Therefore, a sub-optimal design can be constructed by considering alternative simplified expressions for $\text{SINR}_{\text{B}}$ and $\text{SINR}_{\text{E}}$ under certain assumptions. Specifically,  define $\widetilde{\mathbf{R}}$, $\widetilde{\pmb{\sigma}}_{\text{clip}}$, $\widetilde{{\sigma}}^2_{\text{B, norm}}$, and $\widetilde{{{\sigma}}}^2_{\text{E, norm}}$ as the values of $\mathbf{R}$, $\pmb{\sigma}_{\text{clip}}$, ${\sigma}^2_{\text{B, norm}}$, and ${{{\sigma}}}^2_{\text{E, norm}}$, respectively, under the assumption that $\left[\mathbf{v}\right]^2_n +\left[\mathbf{w}\right]^2_n = P_n$  ($\forall n = 1, 2, ..., N_T$). In this case,  $\widetilde{\text{SINR}}_{\text{B}}$ and $\widetilde{\text{SINR}}_{\text{E}}$ represent  $\text{SINR}_{\text{B}}$ and ${\text{SINR}}_{\text{E}}$ when $\mathbf{R} = \widetilde{\mathbf{R}}$, $\pmb{\sigma}_{\text{clip}} = \widetilde{\pmb{\sigma}}_{\text{clip}}$, $\sigma^2_{\text{B, norm}} = \widetilde{\sigma}_{\text{B, norm}}$, and $\widetilde{{\sigma}}^2_{\text{E, norm}} = {\sigma}^2_{\text{E, norm}}$, respectively. We are now focusing on designing $\mathbf{v}$ and $\mathbf{w}$ based on $\widetilde{\text{SINR}}_{\text{B}}$ and $\widetilde{\text{SINR}}_{\text{E}}$. An intuitive reasoning for this sub-optimal design approach is it has been observed that as long as \eqref{constraint21} is satisfied, $\left[\mathbf{v}\right]^2_n + \left[\mathbf{w}\right]^2_n$ tends to achieve its maximum allowable value when \eqref{obj2} is maximized.
Indeed, numerical results in Sec. \ref{result} show that $\widetilde{\text{SINR}}_{\text{B}}$ provides a very good lower bound for ${\text{SINR}}_{\text{B}}$ (less than 5\% difference).

For the sake of analysis, define $\widetilde{\mathbf{h}}_{\text{B}} = \mathbf{h}_{\text{B}}\odot\widetilde{\mathbf{R}}$ and $\widetilde{\mathbf{h}}_{\text{E}} = {\mathbf{h}}_{\text{E}}\odot\widetilde{\mathbf{R}}$. The sub-optimal design problem is then given by
\begin{subequations}
\label{OptProb3}
    \begin{alignat}{2}
        &\underset{\mathbf{v}, \mathbf{w}}{\text{maximize}} & \hspace{2mm} & \widetilde{\text{SINR}}_{\text{B}} \label{obj3}\\
        &\text{subject to }  &  & \nonumber \\
        & & & \widetilde{\text{SINR} }_{\text{E}}  \leq \lambda, \label{constraint31}\\
        & & & \left[\mathbf{v}\right]^2_n + \left[\mathbf{w}\right]^2_n \leq P_n, ~~\forall n = 1, 2, ..., N_T \label{constraint32}.
    \end{alignat}
\end{subequations}
It is easy to see that there exists $\mathbf{v}$ and $\mathbf{w}$ satisfying both \eqref{constraint31} and \eqref{constraint32} regardless of the values of $\lambda$ and $P_n$'s (e.g., $\mathbf{v} = \mathbf{w} = \mathbf{0}$). Hence, \eqref{OptProb3} is always feasible. Also, it is seen that \eqref{OptProb3} is a fractional programming problem, which can often be solved using the Charnes-Cooper transformation \cite{Charnes1962}. Specifically, define
\begin{subequations}
\begin{flalign}
&&\mathbf{v} &= \frac{\overline{\mathbf{v}}}{t},\leftlabel{a} \text{and} & \mathbf{w} &= \frac{\overline{\mathbf{w}}}{t}, \label{b} &&
\end{flalign}
\end{subequations}
where $t > 0$. The problem in \eqref{OptProb3} can then be transformed into the following non-fractional programming problem.
\begin{subequations}
\label{OptProb4}
    \begin{alignat}{2}
        &\underset{\overline{\mathbf{v}}, \overline{\mathbf{w}}, t}{\text{maximize}} & \hspace{0mm} & \left(n_{c}\widetilde{\mathbf{h}}^T_{\text{B}}\overline{\mathbf{v}}\right)^2 \label{obj4}\\
        &\text{subject to }  &  & \nonumber \\
        & & & \left(n_c\widetilde{\mathbf{h}}^T_{\text{B}}\overline{\mathbf{w}}\right)^2 + t^2\left(\left(n_c\mathbf{h}^T_{\text{B}}\widetilde{\pmb{\sigma}}_{\text{clip}}\right)^2 + \widetilde{\sigma}^2_{\text{B, norm}}\right) = 1 \label{constraint41} \\
        & & & \frac{1}{\lambda}\left(n_c\widetilde{\mathbf{h}}^T_{\text{E}}\overline{\mathbf{v}}\right)^2 \leq \left(n_c\widetilde{\mathbf{h}}^T_{\text{E}}\overline{\mathbf{w}}\right)^2 \nonumber \\ & & & \hspace{19mm} + t^2\left(\left(n_c{\mathbf{h}}^T_{\text{E}}\widetilde{\pmb{\sigma}}_{\text{clip}}\right)^2+ \widetilde{\overline{\sigma}}^2_{\text{E, norm}}\right), \label{constraint42}\\
        & & & \left[\overline{\mathbf{v}}\right]^2_n + \left[\overline{\mathbf{w}}\right]^2_n \leq t^2P_n, ~~\forall n = 1, 2, \cdots, N_T. \label{constraint43}
    \end{alignat}
\end{subequations}
From \eqref{constraint41}, we get 
$
    t^2 = \frac{1-\left(n_c\widetilde{\mathbf{h}}^T_{\text{B}}\overline{\mathbf{w}}\right)^2}{\left(n_c\mathbf{h}^T_{\text{B}}\widetilde{\pmb{\sigma}}_{\text{clip}}\right)^2 + \widetilde{\sigma}^2_{\text{B, norm}}}.
$ Substituting  into \eqref{constraint42} and \eqref{constraint43} gives
\begin{subequations}
\label{OptProb5}
    \begin{alignat}{2}
        &\underset{\overline{\mathbf{v}}, \overline{\mathbf{w}}}{\text{maximize}}  \hspace{2mm}  \left(n_{c}\widetilde{\mathbf{h}}^T_{\text{B}}\overline{\mathbf{v}}\right)^2 & &\label{obj5}\\
        &\text{subject to }  &  & \nonumber \\
        & \frac{1}{\lambda}\left(n_c\widetilde{\mathbf{h}}^T_{\text{E}}\overline{\mathbf{v}}\right)^2 \leq \left(n_c\widetilde{\mathbf{h}}^T_{\text{E}}\overline{\mathbf{w}}\right)^2  \nonumber \\ & \hspace{0mm}+ \frac{1-\left(n_c\widetilde{\mathbf{h}}^T_{\text{B}}\overline{\mathbf{w}}\right)^2}{\left(n_c\mathbf{h}^T_{\text{B}}\widetilde{\pmb{\sigma}}_{\text{clip}}\right)^2 + \widetilde{\sigma}^2_{\text{B, norm}}}\left(\left(n_c{\mathbf{h}}^T_{\text{E}}\widetilde{\pmb{\sigma}}_{\text{clip}}\right)^2+ \widetilde{\overline{\sigma}}^2_{\text{E, norm}}\right), & & \label{constraint51}\\
        & \left[\overline{\mathbf{v}}\right]^2_n + \left[\overline{\mathbf{w}}\right]^2_n  \leq \frac{1-\left(n_c\widetilde{\mathbf{h}}^T_{\text{B}}\overline{\mathbf{w}}\right)^2}{\left(n_c\mathbf{h}^T_{\text{B}}\widetilde{\pmb{\sigma}}_{\text{clip}}\right)^2 + \widetilde{\sigma}^2_{\text{B, norm}}}P_n, ~~\forall n = 1, \cdots, N_T. & &\label{constraint52}
    \end{alignat}
\end{subequations}
The problem \eqref{OptProb5} is not convex due to the maximization of a convex function in \eqref{obj5} and the non-convexity of the constraint in \eqref{constraint51}. A well-known technique to solve \eqref{OptProb5} is CCP, which is based on an iterative procedure to find a local optimum \cite{yuille2003,lipp2016variations}. Specifically, at each iteration, the first-order Taylor approximation is utilized to approximately linearlize \eqref{obj5} and the first term in the right-hand side of \eqref{constraint51} as follows.
\begin{align}
    \left(n_{c}\widetilde{\mathbf{h}}^T_{\text{B}}\overline{\mathbf{v}}^{(i)}\right)^2 \approx & \left(n_{c,1}\widetilde{\mathbf{h}}^T_{\text{B}}\overline{\mathbf{v}}^{(i-1)}\right)^2 \nonumber \\ & + 2n^2_{c}\left[\overline{\mathbf{v}}^{(i-1)}\right]^T\widetilde{\mathbf{h}}_{\text{B}}\widetilde{\mathbf{h}}^T_{\text{B}}\left(\overline{\mathbf{v}}^{(i)} - \overline{\mathbf{v}}^{(i-1)} \right),
\end{align}
and
\begin{align}
    \left(n_c\widetilde{\mathbf{h}}^T_{\text{E}}\overline{\mathbf{w}}\right)^2
    \approx & n_c^2\Big(\left(\widetilde{\mathbf{h}}^T_{\text{E}}\overline{\mathbf{w}}^{(i-1)}\right)^2 \nonumber \\ & +  2\left[\widetilde{\mathbf{h}}_{\text{E}}\widetilde{\mathbf{h}}^T_{\text{E}}\overline{\mathbf{w}}^{(i-1)}\right]^T\left(\overline{\mathbf{w}}^{(i)} - \overline{\mathbf{w}}^{(i-1)}\right)\Big),
\end{align}
where $\overline{\mathbf{v}}^{(i)}$ and $\overline{\mathbf{w}}^{(i)}$ are instances of $\overline{\mathbf{v}}$ and $\overline{\mathbf{w}}$ at the $i$-th iteration. Using the above approximations, the CCP involves solving the problem in \eqref{OptProb6} (on top of the following page), which is convex. 
\begin{figure*}[ht]
    \begin{subequations}
\label{OptProb6}
    \begin{alignat}{2}
        &\underset{\overline{\mathbf{v}}^{(i)}, \overline{\mathbf{w}}^{(i)}}{\text{maximize}} & \hspace{1mm} & \left(n_{c}\widetilde{\mathbf{h}}^T_{\text{B}}\overline{\mathbf{v}}^{(i-1)}\right)^2 + 2n^2_{c}\left[\overline{\mathbf{v}}^{(i-1)}\right]^T\widetilde{\mathbf{h}}_{\text{B}}\widetilde{\mathbf{h}}^T_{\text{B}}\left(\overline{\mathbf{v}}^{(i)} - \overline{\mathbf{v}}^{(i-1)} \right) \label{obj6}\\
        &\text{subject to }  &  & \nonumber \\
        & & & \frac{1}{\lambda}\left(n_{c}\widetilde{\mathbf{h}}^T_{\text{E}}\left[\overline{\mathbf{v}}^{(i)}\right]\right)^2 \leq n_c^2\left(\left(\widetilde{\mathbf{h}}^T_{\text{E}}\left[\overline{\mathbf{w}}^{(i-1)}\right]\right)^2  +  2\left[\widetilde{\mathbf{h}}_{\text{E}}\widetilde{\mathbf{h}}^T_{\text{E}}\overline{\mathbf{w}}^{(i-1)}\right]^T\left(\overline{\mathbf{w}}^{(i)} - \overline{\mathbf{w}}^{(i-1)}\right)\right) \nonumber \\  & & & \hspace{32mm} + \frac{1-\left(n_c\widetilde{\mathbf{h}}^T_{\text{B}}\overline{\mathbf{w}}\right)^2}{\left(n_c\mathbf{h}^T_{\text{B}}\widetilde{\pmb{\sigma}}_{\text{clip}}\right)^2 + \widetilde{\sigma}^2_{\text{B, norm}}}\left(\left(n_c\mathbf{h}^T_{\text{E}}\widetilde{\pmb{\sigma}}_{\text{clip}}\right)^2+ \widetilde{\overline{\sigma}}^2_{\text{E, norm}}\right), \label{constraint61}\\
        & & & \left[\overline{\mathbf{v}}^{(i)}\right]_n^2 + \left[\overline{\mathbf{w}}^{(i)}\right]_n^2 \leq \frac{1-\left(n_c\widetilde{\mathbf{h}}^T_{\text{B}}\overline{\mathbf{w}}^{(i)}\right)^2}{\left(n_c\mathbf{h}^T_{\text{B}}\widetilde{\pmb{\sigma}}_{\text{clip}}\right)^2 + \widetilde{\sigma}^2_{\text{B, norm}}}P_n, ~\forall n = 1, \cdots, N_T, \label{constraint62} 
    \end{alignat}
\end{subequations}
\rule{\textwidth}{0.5pt}
\end{figure*}
The detailed procedure is outlined in $\textbf{Algorithm 1}$. 

\begin{algorithm2e}[http]
\SetAlgoLined 
\caption{CCP-type algorithm to solve \eqref{OptProb5}}
\label{alg.3}
Choose the maximum number of iterations $L_{\text{max}}$ and the error tolerance $\epsilon > 0$ for the algorithm. \\
Choose a feasible initial point $\left(\overline{\mathbf{v}}^{(0)}, \overline{\mathbf{w}}^{(0)}\right)$ to \eqref{OptProb6}. \\
Set $i \leftarrow 1$. \\
\While{convergence = \textbf{False} and $i \leq L_{\text{max}}$}{
Solve \eqref{OptProb6} using $\left(\overline{\mathbf{v}}^{(i-1)}, \overline{\mathbf{w}}^{(i-1)}\right)$ obtained from the previous iteration.\\
\eIf{$\frac{\left\lVert\overline{\mathbf{v}}^{(i)} - \overline{\mathbf{v}}^{(i-1)}\right\rVert}{\left\lVert\overline{\mathbf{v}}^{(i)}\right\rVert} \leq \epsilon$ and$\frac{\left\lVert\overline{\mathbf{w}}^{(i)} - \overline{\mathbf{w}}^{(i-1)}\right\rVert}{\left\lVert\overline{\mathbf{w}}^{(i)}\right\rVert} \leq \epsilon$}{
convergence = \textbf{True}. \\
$\overline{\mathbf{v}}^{*} \leftarrow \overline{\mathbf{v}}^{(i)}$. \\
$\overline{\mathbf{w}}^{*} \leftarrow \overline{\mathbf{w}}^{(i)}$. \\
}
{convergence $\leftarrow$ \textbf{False}.\\}
$i \leftarrow i + 1$. \\
}
Return $\overline{\mathbf{v}}^{*}$ and $\overline{\mathbf{w}}^{*}$ then calculate  $\mathbf{v}^{*}$ and $\mathbf{w}^{*}$ from \eqref{a} and \eqref{b}.
\end{algorithm2e} 
The obtained sub-optimal precoders $\mathbf{v}^{*}$ and $\mathbf{w}^{*}$ are used to calculate the actual values of $\text{SINR}_{\text{B}}$,  ${\text{SINR}}_{\text{E}}$ given in \eqref{BobSINR3} and \eqref{EveSINR3}, respectively. For comparison, these values are denoted by $\text{SINR}^*_{\text{B}}$ and ${\text{SINR}}^*_{\text{E}}$ while the respective values obtained
though the approximation of Sec.~\ref{sec:ANdesign}
are denoted by $\widetilde{\text{SINR}}^{*}_{\text{B}}$ and $\widetilde{{\text{SINR}}}^{*}_{\text{E}}$. Similarly, $\text{SINR}^*_{\text{B, NoAN}}$, ${\text{SINR}}^*_{\text{E, NoAN}}$, $\widetilde{\text{SINR}}^{*}_{\text{B, NoAN}}$, and $\widetilde{{\text{SINR}}}^{*}_{\text{E, NoAN}}$ are defined in the same manner for the case of the no-AN scheme (i.e., $\mathbf{w}=0$). 

\section{Numerical Results and Discussions}
\label{result}
This section presents simulation results to demonstrate the secrecy performance of the proposed AN designs. The system configuration and parameters are given in Table I, and simulation results are obtained by averaging over 5000 randomly located positions of Bob and Eve. 
\begin{table}[ht]
\caption{System Parameters} 
\centering 
\begin{tabular}{l l l} 
\midrule\midrule
Room dimension, &  5 $\times$ 5 $\times$ 3 \\ Length (m) $\times$ Width (m) $\times$ Height (m) \\
 \midrule
 LED luminary positions & Luminary 1 : ($-\sqrt{2}$, $-\sqrt{2}$, 3) \\ & Luminary 2 : ($\sqrt{2}$, $-\sqrt{2}$, 3)\\ & Luminary 3 : ($\sqrt{2}$, $\sqrt{2}$, 3) \\ & Luminary 4 : ($-\sqrt{2}$, $\sqrt{2}$, 3) \\
 \midrule     
 LED bandwidth, $B_{\text{mod}}$ & 20 MHz \\
 \midrule
 Number of chips in each luminary, $n_c$ & 24 \\
 \midrule     
 LED beam angle, $\phi$ & $120^\circ$ \\ (LED Lambertian order is 1) \\
 \midrule 
 LED conversion factor, $\eta$ & 0.44 W/A  \\
 \midrule
 $[I_{\text{min}},~~I_{\text{max}}]$ & $[0\text{A},~~1\text{A}]$ \\
 \midrule \midrule    
 Active area, $A_r$ & 1 $\text{cm}^2$ \\ 
 \midrule     
 Responsivity, $\gamma$ & 0.54 A/W\\ 
 \midrule     
 Field of view (FoV), $\Psi$ & $70^\circ$\\ 
 \midrule     
 Ambient light photocurrent, $\chi_{\text{amp}}$ & 10.93 $\text{A}/(\text{m}^2 \cdot \text{Sr}$)\\
\midrule 
Preamplifier noise current density, $i_{\text{amp}}$ & 5 $\text{pA}/\text{Hz}^{-1/2}$ \\
\midrule \midrule
\end{tabular}
\label{table1}
\end{table} 

\begin{figure}[t]
    \centering
    \includegraphics[width = .5\textwidth, height = 6.2cm]{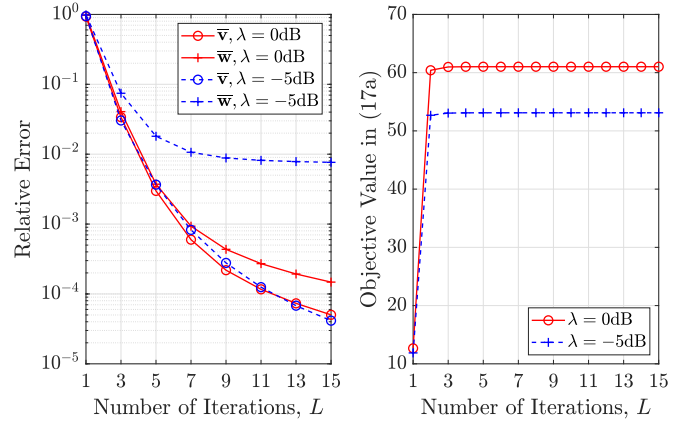}
    \caption{Convergence behaviors of $\textbf{Algorithm 1}$.}
    \label{convergenceAlgo1}
\end{figure}

Firstly, Fig.~\ref{convergenceAlgo1} depicts the convergence behavior of  $\textbf{Algorithm 1}$ with respect to the number of iterations $L$ when $\lambda = 0$ and $-5$dB. It is shown that, on average, the objective function in \eqref{obj6} approaches its optimal value when $L\approx 3$, which corresponds to relative errors of $\overline{\mathbf{v}}$ and $\overline{\mathbf{w}}$ being greater than $10^{-2}$, respectively. Accordingly, the error tolerance $\epsilon = 10^{-2}$ and the maximum number of iteration $L_{\text{max}} = 10$ can be chosen to guarantee a satisfactory convergence of \eqref{obj6}. 
\begin{figure*}[t]
    \begin{subfigure}[b]{\textwidth}
        \centering
        \includegraphics[width = .7\textwidth, height = 4.55cm]{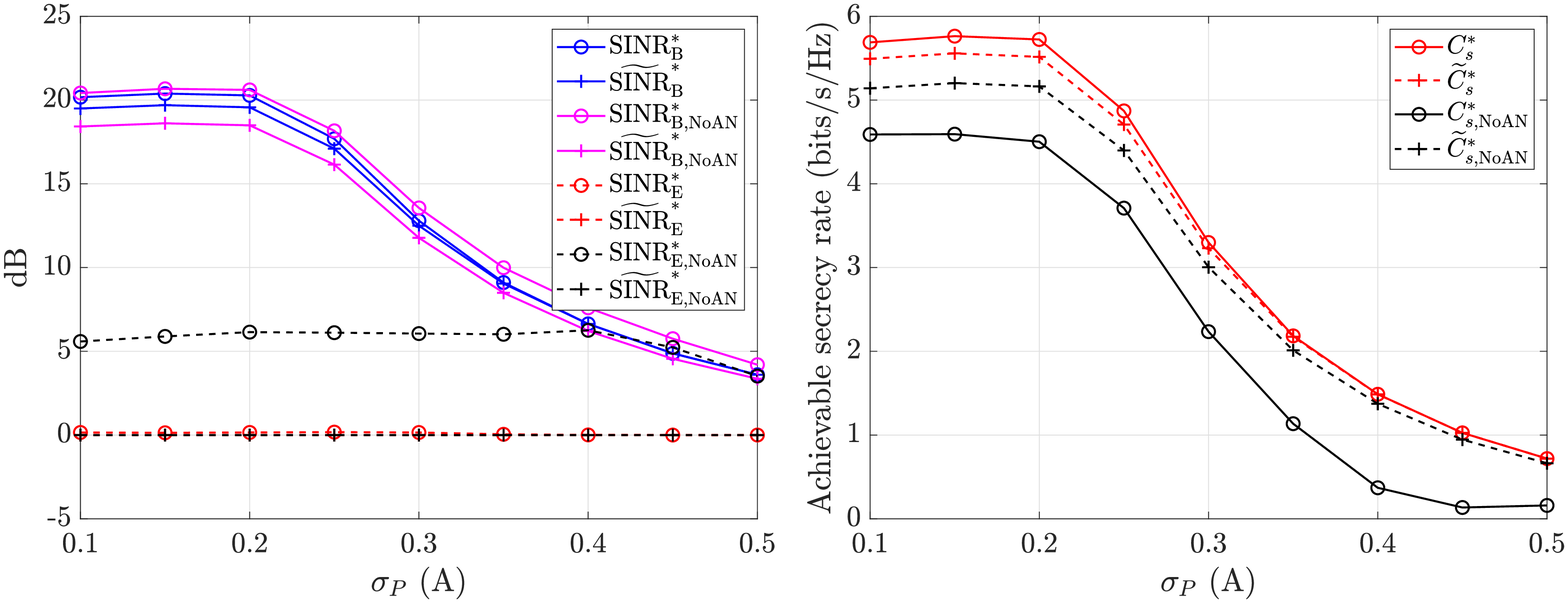}
        \caption{$\lambda = 0$ dB.}
        \label{SecrecyPower1}
    \end{subfigure}
    \begin{subfigure}[b]{\textwidth}
        \centering
        \includegraphics[width = .7\textwidth, height = 4.55cm]{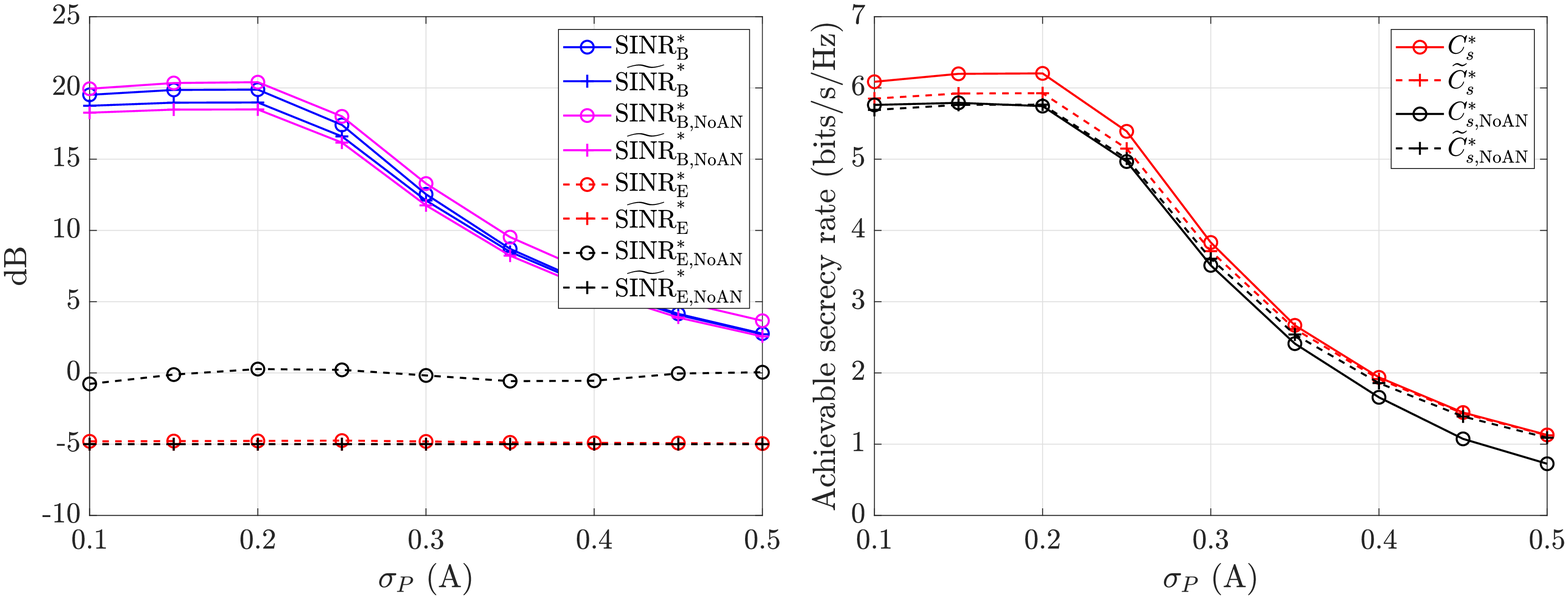}
        \caption{$\lambda = -5$ dB.}
        \label{SecrecyPower2}
    \end{subfigure}
    \caption{Secrecy performance versus $\sigma_{P}$ for different thresholds on Eve's SINR.}
    \label{Secrecy_vs_Power}
\end{figure*}

Assuming DCO-OFDM is employed to generate the information-bearing signal and AN, a widely used setting for $I_{\text{DC}}$ is $I^2_{\text{DC}} = 4P_n$ is employed, which is equivalent to 7dB bias \cite{Armstrong2008}. If we denote $\sigma_{P_n} = \sqrt{P_n}$ (i.e., $\sigma_{P_n}$ represents the maximum allowable standard deviation for the combined information-bearing and AN signals), then $I_{\text{DC}} = 2\sigma_{P_n}$. Consider that $\sigma_{P_n}\text{'s} = \sigma_{P}$, Fig.~\ref{SecrecyPower1}  and Fig.~\ref{SecrecyPower2} present the secrecy performance as a function of $\sigma_{P}$ for $\lambda = 0$ dB and $-5$ dB, respectively.  Firstly, the left sub-figures compare the AN and no-approaches in terms of Bob’s and Eve’ SINR performances. In the AN scheme, notice that the difference between $\text{SINR}^{*}_{\text{B}}$ and $\widetilde{\text{SINR}}^*_{\text{B}}$ are small, especially at high values of $I_\text{DC}$. More
importantly, the differences between $\text{SINR}^*_{\text{E}}$ 
and $\widetilde{\text{SINR}}^{*}_{\text{E}}$ are seen to be negligible. These two observations confirm the approximation of the proposed sub-optimal approach presented in Sec.~II-B. In the case of no-AN scheme, the gap between $\text{SINR}^*_{\text{B, NoAN}}$ and $\widetilde{\text{SINR}}^*_{\text{B, NoAN}}$ are more clear but still near 3 dB. Nonetheless, there are considerable gaps between $\text{SINR}^*_{\text{E, NoAN}}$ and $\widetilde{\text{SINR}}^*_{\text{E, NoAN}}$, where Eve’s SINRs are about 5 dB higher than the results obtained from using the approximate expression.

For a better illustration of the secrecy performance, the right sub-figures compare the AN and no-AN schemes in terms of achievable secrecy rate. Since the information-bearing and AN signals are approximated to be Gaussian, the achievable secrecy rates corresponding to the pairs $\left(\text{SINR}^*_{\text{B}},~\text{SINR}^*_{\text{E}}\right)$ and $\left(\widetilde{\text{SINR}}^*_{\text{B}}, ~\widetilde{\text{SINR}}^*_{\text{E}}\right)$ in the case of known $\mathbf{h}_{\text{E}}$ are given by $C^*_s = \log_2\left(1 + \text{SINR}^*_{\text{B}}\right) - \log_2\left(1 + {\text{SINR}}^*_{\text{E}}\right)$ and $\widetilde{C}^*_s = \log_2\left(1 + \widetilde{\text{SINR}}^*_{\text{B}}\right) - \log_2\left(1 + \widetilde{{\text{SINR}}}^*_{\text{E}}\right)$, respectively. Note that $C^*_{s, \text{NoAN}}$ and $\widetilde{C}^*_{s, \text{NoAN}}$ are similarly defined. Then, it is clear to see the superiority of the AN scheme over the no-AN approach. For example, at $\sigma_{P} = 0.25A$, the use of AN improves the achievable secrecy rate by about $1.2$ and $0.4$ bits/s/Hz when $\lambda = 0$ and $-5$ dB, respectively. Moreover, observe that at low $\sigma_{P}$ values (i.e., below $0.2$ A), the achievable secrecy rate stays nearly unchanged. This arises because, for low $\sigma_{P}$, the impact of increasing $\sigma_{P}$ is balanced by the clipping
distortion. However, when $\sigma_{P} > 0.2$ A, the impact of the clipping distortion becomes dominant, resulting in a dramatic degradation of the achievable secrecy rate.
\section{Conclusion}
\label{conclusion}
In this paper, we have studied an AN design for VLC systems considering the impact of  clipping distortion. Simulation results revealed that compared with the no-AN scheme, the proposed design could effectively constrain Eve's SINR to the predefined threshold while did not affect considerably Bob's SINR. As a result, using AN was beneficial in improving the secrecy performance. Furthermore, significant impacts of the clipping distortion on the secrecy performance were also observed. Our future work will focus on novel signal transmission schemes to alleviate the clipping distortion, AN design for the case of multi-user, and examine the case that Eve's CSI is unknown at the transmitters. 
\bibliographystyle{IEEEtran}
\bibliography{references}
\end{document}